# Is 3D chip technology the next growth engine for performance improvement?

P. G. Emma
E. Kursun

*The semiconductor industry is reaching a fascinating confluence in several evolutionary trends that will likely lead to a number of revolutionary changes in how computer systems are designed, implemented, scaled, and used. Since Moore's Law, which has driven the evolution in systems for the last several decades, is imminently approaching real and severe limitations, the ability to create three-dimensional (3D) device stacks appears promising as a way to continue to integrate more devices into a "chip." While on the one hand, this nascent ability to make "3D technology" can be interpreted as merely an extension of Moore's Law, on the other hand, the fact that systems can now be integrated across multiple planes poses some novel opportunities, as well as serious challenges and questions. In this paper, we explore these various challenges and opportunities and discuss structures and systems that are likely to be facilitated by 3D technology. We also describe the ways in which these systems are likely to change. Since 3D technology offers some different value propositions, we expect that some of the most important ways in which 3D technology will likely impact our approach to future systems design, implementation, and usage are not yet obvious to most system designers, and we outline several of them.*

## Introduction and motivation

In the past decade, there have been numerous process demonstrations that illustrated that two or more chips can be joined in various ways through stacking, thus utilizing the third dimension. While similar forms of this technology have already been deployed in DRAM in order to increase the capacity of dual inline memory modules (DIMMs), there are significant efforts to import the three-dimensional (3D) stacking technology to the logic domain. Often, the immediate intuitive reaction to the concept of 3D integration is that it must obviously be enormously useful in general, but there will be a dramatic temperature increase in the components. Both of these instinctive reactions are partly true, but neither is true absolutely. In this paper, we discuss the numerous advantages and challenges associated with potential uses for 3D technology.

Interestingly, 3D technology is emerging as numerous other paradigms in silicon technology and system design are shifting. Device speed is not scaling below the 65-nm node the way it had historically scaled. Furthermore, leakage currents appear to be a daunting impediment to progress below the 45-nm node. Lithographic solutions beyond 32 nm still pose significant uncertainty. Device variability is also becoming a problem as planar dimensions and dopant levels decrease. The increase in device variability requires adding large margins to the circuit-level specifications in order to ensure correct operation as well as reducing the overall yield. In short, Moore's Law appears to be slowing greatly, if not coming to an end. The question is whether 3D technology will be the next scaling engine for semiconductor technology.

Three-dimensional integration might alleviate a number of immediate problems faced by system architects. The first problem is the infamous "memory wall." Historically, processor performance has improved by about 60% per year, whereas the corresponding improvement in memory access time has been less than

541

10% per year. This gap is a major factor that limits overall system performance improvement and is just one of several factors comprising the memory wall. This effect can be even more pronounced with the increasing number of cores in the multicore context, which we elaborate in the section on concurrent trends in systems.

The potential to improve the interconnect latency using 3D technology to ease the effects of the gap between processor performance and memory access time appears promising. Three-dimensional technology can enable the integration of memory layers onto the processor chip and can thereby eliminate the slower and higher-power off-chip buses to that memory by replacing them with high-bandwidth and low-latency vertical interconnections to the memory layers [1]. This kind of configuration not only increases the on-chip cache memory capacity dramatically but also improves the access latency at the same time. Logic and memory chips that require tens of millimeters per wire to connect in the planar domain can instead be interconnected using through-silicon vias (TSVs) that are merely tens of microns long in the vertical dimension, which roughly translates to a three orders of magnitude difference. The wire-limited performance improvement through vertical integration is projected to be the square root of the number of layers in a 3D stack [2]. Other studies have also shown that increasing the number of active layers through vertical integration improves the interconnection performance and bandwidth significantly [3, 4].

Memory latency scaling is not the only issue that designers have to overcome. Next-generation processors with tens of cores and tens of billions of transistors are challenging the industry with a wide range of issues including packaging, interconnect, power delivery, thermal management, and design complexity. Even though the complexity and functionality of microprocessors had been increasing, it is likely that many of these trends will change into new scaling trends as the difficulties in traditional silicon scaling increase and as the value propositions in system scaling change. Integration of 3D provides solutions for several scaling issues such as reducing the areal footprints of system components by increasing the volumetric density of transistors by way of stacking chips.

Footprint or size reduction has been the main driving factor for adoption of 3D technology in embedded systems. However, for microprocessors, the ability to increase bandwidth will become a limitation as processor performance increases. At the same time, chip areas have been growing steadily. For example, the Intel Quad-Core Itanium** is reported to have a footprint area of more than 700 mm$^2$ [5]. Despite all design efforts, the corresponding increase in the area causes chip-crossing latencies to increase, and it compounds yield issues [6].

The opportunity for mixed-signal or mixed-technology designs (analog, memory, RF, field-programmable gate array [FPGA], CMOS, SiGe) creates novel design possibilities in vertical integration. Use of 3D enables heterogeneous integration of disparate signals and technologies effectively within the same stack, where each layer is processed in the target process technology without being affected by common manufacturing complications in the traditional two-dimensional (2D) manufacturing counterpart. Such heterogeneity can be leveraged to incorporate advanced security and reconfigurability features, as well as specialized accelerators for performance improvement in traditional microprocessor architectures.

Manufacturing and testing complications as well as power and thermal issues are considered to be among the major challenges of vertical integration. The on-chip temperature profile increases considerably in 3D logic stacking because of the increased power density on a smaller footprint and higher thermal resistances. Current research indicates that with the recent advances in thermal vias and in 3D manufacturing processes, thermal problems are expected to be difficult, though not insurmountable [7]. We elaborate more on these problems and explore some proposed solutions in later sections of this paper.

## 3D integration technology: Preliminaries

Early 3D work started as pure technology studies in the late 1970s [8] and 1980s [9–11]. Following this trend, and continuing through most of the 1990s, 3D technology remained mostly a research concept. Until the turn of the century, commercial electronics had primarily been limited by logic speed, and not by interconnections. Moreover, early versions of vertical integration did not offer great interconnectivity between layers—for example, only 200–250 vertical interconnects in a 3D stack [10]. Until recently, the practical interconnectivity within chip stacks was mostly limited to wire bonding at the periphery, which severely limits the number of connections. Recent improvements in TSVs enable more than 100,000 vias per square centimeter and result in a wide range of applications where the interconnection lengths are reduced from tens of millimeters to tens of microns [12].

Today, a large number of academic groups and companies are involved in 3D research. Vertically integrated chips have become available commercially in embedded, wireless, and memory devices. Systems in these sectors have strict area and volume limitations. Hence, shrinking the chip footprint delivers high value in the corresponding markets. The microprocessor sector will likely be next to adopt 3D integration, but the value proposition is clearly different, with the focus on improved interconnectivity, packaging density, memory latency, and bandwidth improvement as well as heterogeneous integration.

Three-dimensional technology spans a wide range of techniques, with significant variation in the corresponding characteristics. The manufacturing technology includes the sequential silicon growth approach at the exploratory end of the spectrum [13, 14] as well as parallel integration at the wafer or chip level where preprocessed wafers or dies are integrated in a stack through TSV formation (as shown in **Figure 1**). These techniques vary in terms of their baseline characteristics such as interlayer interconnect lengths and bandwidth, and therefore, in the architectural features that they enable. References [15–23] provide a good overview of the current 3D manufacturing techniques. We focus on wafer- and chip-level parallel integration for the rest of the discussion.

### Silicon-carrier technology

Integration of multiple chips on a shared carrier—such as a processor with a DRAM chip—significantly reduces the interconnection delay due to the lower capacitances, older interconnections, and greatly reduced interconnect distances. As shown in **Figures 2(a)** through **2(c)**, *silicon carrier* generally refers to a silicon substrate that incorporates 1) a dense I/O grid with high-speed standard back-end wiring on one side; 2) coarser-grain controlled-collapse chip connection (C4) solder bumps on the other side to facilitate connection to an organic or ceramic package; and 3) a vertical TSV infrastructure that connects the two sides together. Microbump technology can be used on the wiring surface of the silicon carrier to connect two or more chips to the carrier. Figure 2(a) is a photograph of four silicon chips shown mounted on two silicon-carrier packages on a glass–ceramic substrate. Figure 2(b) is a higher magnification view showing the edge of one of the silicon carriers [24].

Microbumps with 50-$\mu$m diameters on 100-$\mu$m centers were demonstrated by Patel et al. in 2005 [25]. State-of-the-art 3D technology enables 25-$\mu$m-diameter microbumps on 50-$\mu$m centers; such technology can also be used to enhance power distribution and the I/O infrastructure of a 3D stack. Up to 16 times higher density over standard chip I/O was demonstrated in recent studies [26, 27] as a result of the improvement in microbump pitch.

### 3D stacking

**Figure 2(d)** illustrates a 3D stack, which can be in a face-to-face (F2F) position, with device layers facing each other, or a face-to-back (F2B) position, with device layers both facing in the same direction. Note that when integrating more than two layers, an F2B process (or

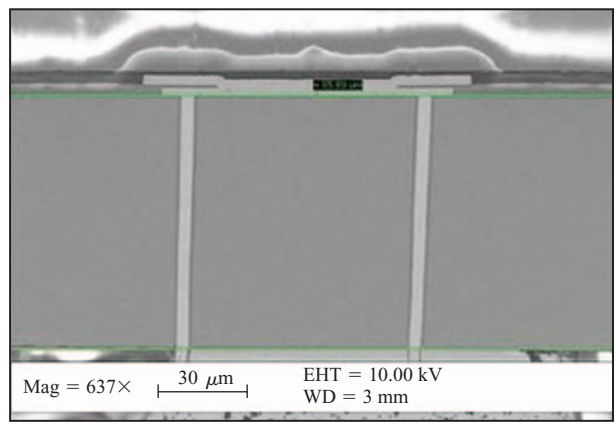

Mag = 637×    30 $\mu$m    EHT = 10.00 kV
                          WD = 3 mm

### Figure 1

Cross-sectional micrograph showing two through-silicon vias. (EHT: electron high tension; WD: working distance.)

B2B) is needed in the stack, although the stack might terminate with an F2F layer in order to facilitate connection to a package, depending on the processes used. TSVs are used for interconnecting the device layers in the wafer or for chip-level integration.

Even though F2F integration refers to the immediate interconnectivity of top-metal layers of the corresponding chips, TSVs are still required for power delivery and other I/O connections. Hence, F2F bonding provides a higher signal interconnection density than alternative bonding techniques such as F2B [28, 29], which rely purely on TSVs for all signal and I/O interconnections. The current sizes of TSVs range from 1 to 10 $\mu$m, depending on the characteristics of the individual 3D manufacturing process [30].

Although wafer-bonding techniques vary in their process flow, they share common 3D-specific manufacturing stages such as 1) silicon substrate thinning, 2) wafer alignment, 3) bonding, and 4) vertical interconnect. A major part of the 3D integration process flow follows the front-end processing in wafer- or die-level integration: Individual wafers are bonded following the thinning and polishing of the surface for better bonding quality. Subsequently, wafers are bonded to each other by using a metal-to-metal thermocompression process or by using polymeric or dielectric glue layers. The TSV interconnection stage can potentially happen in earlier or later stages of the manufacturing process.

Functional verification of each layer can be completed before the bonding in such wafer- or chip-level integration, which improves the overall yield. The bonding of multiple wafers creates complications as a result of wafer-to-wafer alignment and reliability issues

**543**

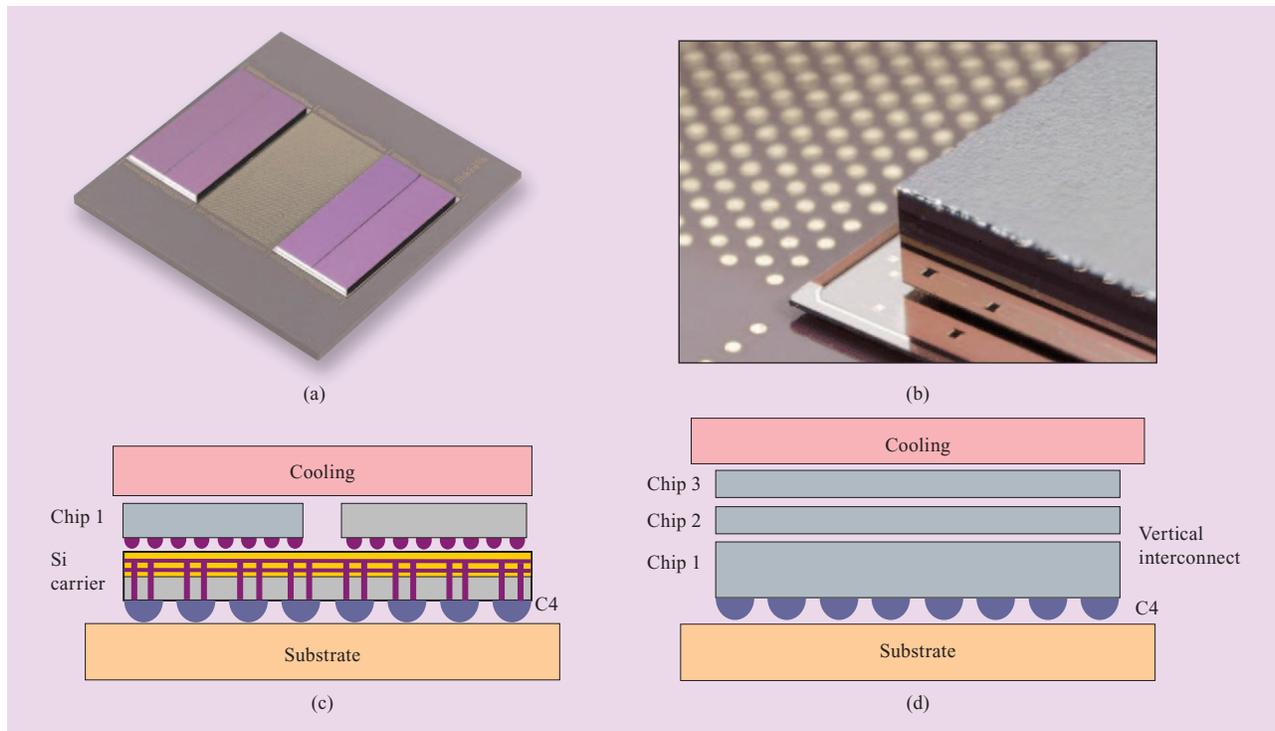

arising from the high-temperature and pressure-based bonding techniques.

Since device layers are fabricated separately, the wafer- or die-level parallel integration is especially promising for heterogeneous integration of disparate technologies (analog, digital, RF, silicon-on-insulator [SOI], SiGe) within the same stack. It is very important to note that there are a number of variations for each technique, and in many cases, characteristics of various implementations and manufacturing process flow may be significantly different. Further details of the individual techniques can be found in the listed references.

## Concurrent trends in systems

In this section, we discuss a number of architectural trends and the potential implications of 3D system architectures. Through the 1990s, and into the turn of the century, the trends in processor design were to drive the frequency upward and to make more complicated microarchitectures to extract as much instruction-level parallelism (ILP) as possible from a running program. As improvements in device speed have slowed, and as the limitations of power have become the dominant factor in systems evolution, an industry-wide "cease fire" has been declared when it comes to frequency scaling. It appears unlikely that the industry will try to push processor frequency beyond where it is today (in the single digits of gigahertz) due to power efficiency considerations.

Similarly, extracting ILP has also faced diminishing returns, because much of the speculation required to do this is wasted computation, and because the extra hardware required for parallel computation consumes power when it is not needed. Since leakage is becoming a significant factor in power dissipation, it is even more difficult to justify superscalar[1] designs. The trend toward wide-issue designs that took off in the 1990s has waned. It is clear that future scaling will not rely on driving the frequency up, and it will not rely on aggressively extracting ILP. However, there are three new trends in systems scaling that increase system performance in new and possibly ultimately limiting ways. Specifically, industry has been cramming more and more processing threads (and hence, working sets) into computing nodes

---

[1]*Superscalar systems* execute multiple instructions per cycle.

and chips via three new trends, each stemming from a different set of motivations.

The IBM pSeries* G4 system (2001) introduced the multicore concept with chips having two processing cores, which doubles the computation capacity of a chip. Since the chip supported two programs (threads) running simultaneously—one on each core—it also required nearly twice the off-chip bandwidth and twice the on-chip cache capacity of a single-core chip. Depending on the workload, the requirements can actually be greater than these.

The multicore notion is depicted in **Figure 3(a)**, where the processor chip is shown to contain four processing cores (P) and their associated levels of cache (L1, L1.5, and L2) as well as the nest, which comprises the interconnection fabric and the I/O and memory controllers. Note that as cores are added to a chip, an increasing percentage of the chip area is needed for the cache capacity required to support all of the additional computation. What is not depicted here is that the off-chip bandwidth required to support this additional computation must increase as well. There is a nonlinear relationship between storage capacity and bandwidth that can be quite detrimental, as explained in the next section.

The IBM pSeries G5 system further raised performance potential by adding concurrent threads with multithreading technology. In the G5, each core is capable of holding two independent copies of an architected state[2] so that two programs could run simultaneously on each core and share the functional resources of those cores in order to increase their utilization; that is, each core behaves as if it were two independent cores, which puts further pressure on the on-chip caches and off-chip bandwidth as depicted in **Figure 3(b)**. Each of the cores shown in Figure 3(a) is shown as having four color stripes, with each stripe denoting a different concurrent thread. Therefore, the four-core chip shown in Figure 3(b) appears to be a 16-core (or at least, a 16-thread) chip. The software layer cannot distinguish these cases. It should be clear that this puts even more pressure on the on-chip cache and on the off-chip bandwidth.

Ironically, the reason that IBM began to use multithreading was to improve the utilization of all of the lightly utilized parallel components of the ancestral microarchitectures built for high ILP. Light average utilization of resources is a common characteristic of a high-ILP engine. This is one of the reasons that high-ILP designs have disproportionately high leakage currents and are, thus, inefficient. Those high-ILP ancestral microarchitectures were designed to provide fast turnaround on single-threaded applications. Instead,

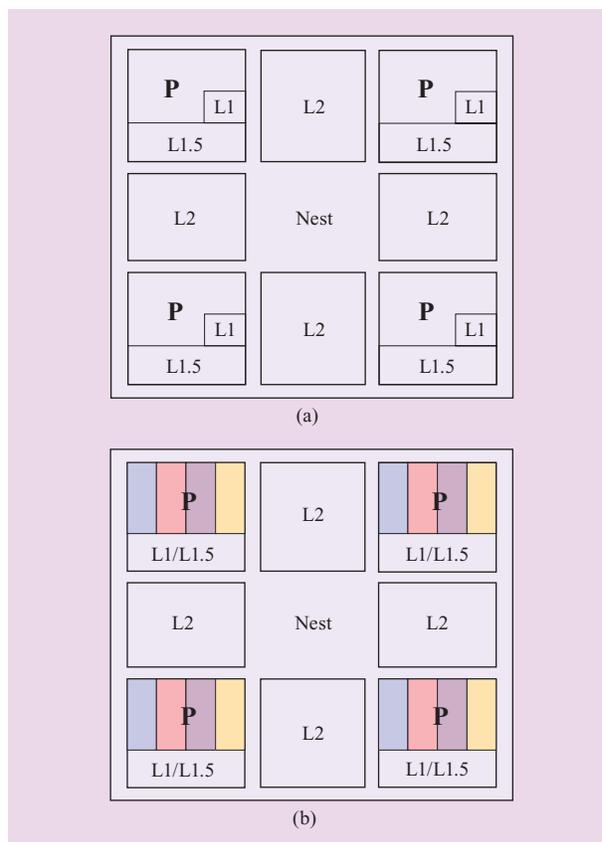

multithreading provides high throughput for many threads.

The third trend is the wider availability of virtualization technology within the software stack to create more simultaneous and co-resident system images (i.e., on the same hardware), and the impending desire to use virtualization technology at lower levels in the software stack, such as the application layer. Basically, virtualization technology gives the software stack the ability to create a virtual machine on which an independent operating system can run. It can then create many independent virtual machines in order to provide independent (hence, secure) systems for many different users and applications. While this is advantageous from a security and robustness perspective, all of these virtual systems can be co-resident on the same set of real cores and caches. **Figure 4** illustrates the modified version of the chip from Figure 3(b) with four independent and co-resident threads running on it. From the outside, this chip now appears to be four independent 16-core systems.

---

[2]The *architected state* is that part of the operating state which is explicitly visible to the running program.

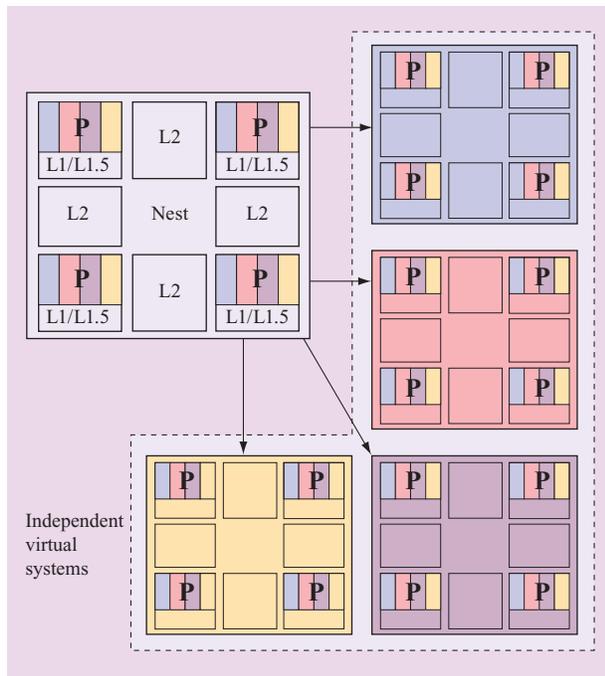

Obviously, this compounds the stress placed on the on-chip caches and on the off-chip bandwidth.

### Cache capacity and bandwidth

It is known heuristically that the cache-miss rate is proportional to the reciprocal of some root of the cache capacity. Recently, it was shown that this is because the probability of re-references to data is driven by a power law [31]. For many workloads, the square root is a good fit. Data in a cache is stored as *cache lines*, which are contiguous sequences of bytes, aligned on boundaries of some power of 2, for example, 64 bytes or 128 bytes, or even more. The reason for this is that reference patterns follow two orthogonal localities, so large chunks (cache lines) are used to capture these behaviors.

Spatial locality of reference is the phenomenon that if a program references a particular datum, then it is extremely likely that the program will also reference other data that are spatially close (by address) to the referenced datum; that is, it will reference *nearby data*. This is the rationale for bringing in 64 or 128 bytes instead of merely words (4 bytes). Misses caused by near-future references to nearby data (within 64 or 128 bytes) are obviated. Temporal locality of reference is the phenomenon that if a program references a particular datum, then it is extremely likely to re-reference that same datum again in the near future. This locality is the rationale for caching data in the first place; if it were not inherent to programs, caches would not work. Most real workloads exhibit both behaviors simultaneously.

Choosing the right line size in a cache is then an optimization problem in which spatial and temporal localities are traded against each other within a fixed-capacity cache. If a cache is partitioned into many short lines, the fact that there are many of them will capitalize on the temporal aspects of the reference behavior. If the cache is partitioned instead into a smaller number of larger lines, then less temporal context can be captured, but the longer lines provide more spatial context.

However, note that there are other effects of the choice of line size. A cache directory must have a directory entry per line. If the size of the cache is doubled, then the choice is either to keep the size of the directory (therefore, its tractability) intact by doubling the line size or to keep the line size intact and double the size of the directory. Since directory access time is considered sacrosanct in many machines, the trend is toward larger line sizes as cache sizes grow. (In reality, cache line sizes are not generally growing in commercial systems because of contention and potential software tuning problems. The argument just describes one of the incentives to grow the line size.)

An adverse effect of doubling the line size is that it will then take twice as long to transfer a line if the bandwidth is held constant. This can cause costly queuing delays if certain bus utilization thresholds are exceeded, since queuing delay has a severe nonlinearity as utilization increases. The time to transfer a line (measured in processor cycles) is called the "trailing edge" (TE) of a miss and is equal to the number of packets in the line (which is the line size divided by the bus width) times the transfer rate (processor clocks per bus clock). Note that the bus utilization is proportional to the TE, since from the perspective of the bus, the TE is the service time.

In fact, when measuring the performance of a system, the bus bandwidth and the cache capacity manifest as each other and are "mutually fungible" in the following obvious way. To the extent that a cache can be made larger, less bandwidth is required to sustain the cache contents, since those contents will persist longer in the cache. In addition, to the extent that the bandwidth can be increased, less cache capacity is required, since the bandwidth enables the cache to be more facile at quickly importing its contents.

However, because of the power law that governs the miss rate as a function of capacity, this "fungibility" between bandwidth and capacity is disproportionate, as we show below. The reason that this is an impending problem is that we are nearing reasonable electrical off-chip bandwidth limits (and these are area and power limits) and will likely hit them in a generation or two. Further down the road, optics has the potential to provide some relief, but it will likely not arrive in time to avoid the problem.

If we use $T$ to represent performance (where $T$ is the number of threads at a fixed speed), $B$ to represent off-chip bandwidth, and $C$ to represent cache capacity, the impending question will be, "If $B$ cannot increase, what must we do to $C$ to be able to double $T$ (e.g., by adding cores, or threads, or virtual images, which is clearly the trend)?" **Figure 5** shows the situation with $T$ threads, $B$ bandwidth, and $C$ cache.

If we double $T$ by merely putting two copies of this system on the same chip to get $2T$ (see **Figure 6**), then we also double $C$ (to get $2C$), and we double $B$ (to get $2B$). However, if we are at a bandwidth limit, we must hold $B$ fixed. This means that we need to cut the original $B$ in half. Because the miss rate varies as the square root of the cache capacity, cutting $B$ in half requires quadrupling $C$ for each $\{T, B, C\}$, in order to get two copies of $\{T, B/2, 4C\}$.

This means that when we are bandwidth limited, doubling $T$ requires scaling the cache capacity by a factor of 8 (i.e., by $2 \times 4C$). Since the apparent trends in future scaling all involve increasing the number of threads, and since we are nearing bandwidth limits, this implies that we need a technology that can integrate more storage at an extremely high rate of scaling. This is quite formidable and beyond the capacity of technology that follows Moore's Law.

## Composing systems in 3D

Assuming that 3D integration is a viable technology for providing the high scaling implied by the trends discussed above, the next natural question to ask is how to best compose 3D systems. A basic approach is stacking existing 2D chips into a 3D system having correspondingly more cores and storage, as shown in Figure 6(a). The appeal of this approach is that it uses existing systems as building blocks with a minimum amount of change in the design. In fact, it resembles the design of blade systems taken down to the chip level.

Conspicuously missing from this approach is that the system was not "conceived" as a 3D system, that is, no special optimization or design effort was dedicated to taking advantage of the third dimension. As a result, the system does not utilize the full potential of 3D capabilities. For instance, if the original 2D chip layouts are not optimized for 3D, then hotspot locations overlap in the vertical dimension, worsening the heating problem. In other words, in a multistrata system, the original hotspots will become severe hotspots, and the thermal gradients (when viewed from a planar perspective) will be severe. Moreover, it has been shown that the design stages from architecture layer to the physical design need to be optimized and integrated, specifically for 3D, in order to achieve a significant benefit [32].

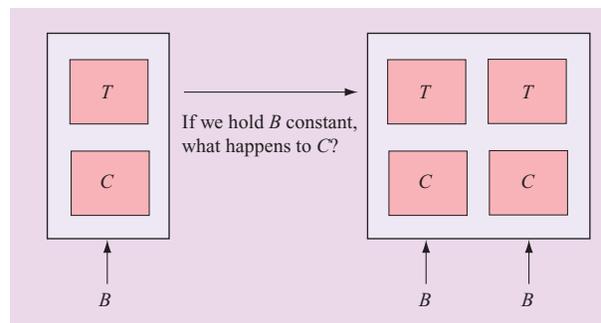

A second approach is to partition the components of the system onto different layers on the basis of functionality. Figure 6(b) illustrates a possible implementation with the first device layer consisting of processing cores and minimum-size cache thereby forming a processor layer, above which could be cache layers or layers providing other functionality (I/O infrastructure, test infrastructure, accelerator infrastructure, etc.). This has the advantage that the system becomes "componentized" into subsystems that are independently testable prior to assembly. The system components can further be optimized at different technology nodes, such as older technologies for cost optimization, with favorable interconnectivity requirements. Furthermore, these subsystems are more thermally homogeneous than their current planar counterparts; hence, the resulting stack should not be plagued by severe thermal gradients. Thermal gradients become more prominent for the thinned silicon device layers in the 3D stack [33].

Finally, with the availability of high-bandwidth interlayer interconnect, a third approach may emerge: to make spherical macros and processors. By "spherical," we imply spreading the functionality across multiple adjacent strata at a finer granularity. This has natural appeal because the radius required to contain a fixed number of gates in 3D space is shorter than it is in 2D space, hence there is a timing advantage to building spherical processors. Even though most of the performance improvement in wire-length reduction through vertical integration is observed in global wiring, local or intrablock wiring can also provide further improvement potential [34]. Note, however, that in current technology nodes, device variability is becoming a

**547**

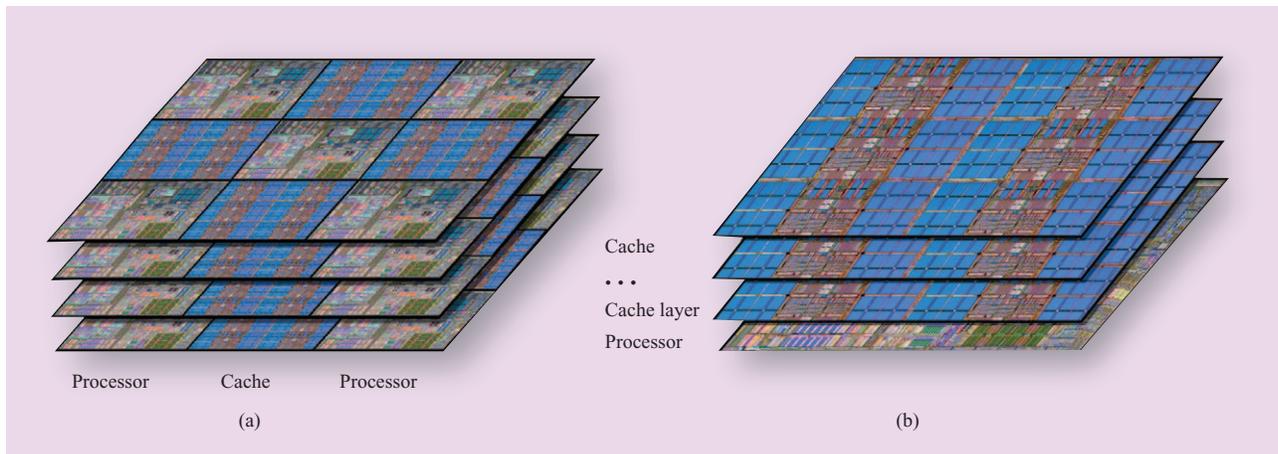

Cache
• • •
Cache layer
Processor

Processor          Cache          Processor

(a)                                    (b)

**Figure 6**

Potential 3D systems: (a) a 3D system built from 2D systems; (b) allocation of layers by functionality.

huge problem within a single chip. The degree to which variability is further amplified when driving signals across multiple strata is unknown and could negate the benefits of the shorter radius in a sphere for wafer- and chip-level integration alternatives. Furthermore, finer grain integration requires strong 3D electronic design automation support because of the extended design space.

## Unique advantages and challenges of 3D systems

### Advantages
One of the basic advantages of 3D structures is the improved volumetric density as well as allowing the connections between the components to be shorter. This, in turn, allows those connections to be faster and more power efficient. At the same time, 3D technology provides improved interconnectivity with the increased number of neighbors in both lateral and vertical dimensions. Modern microprocessor architectures are under constant pressure to add more metal layers. Wire- and port-limited blocks are greatly affected by this limitation and can potentially benefit more from 3D integration than logic-limited counterparts. Yet, routing data in three dimensions can be efficient, especially if functional elements have been arranged within the stack so that the connections between them are mostly vertical. In this case, the routing is merely done in the $z$ dimension, with smaller $x$ or $y$ displacement. Moreover, in the special case of buses between cache layers, the bandwidths can be made truly massive if the bits within those cache layers are placed with a little forethought. In fact, problems associated with long lines and long trailing edges can be nearly eliminated.

There are two obvious optimizations that can be made in the physical placement of bits within cache layers in order to minimize the $x$–$y$ dislocations when moving cache lines between caches in a hierarchy. For example, a cache line of 128 bytes has 1,024 bits plus parity or error correcting code (ECC) bits. The first observation is that when moving a cache line from one cache level to another, bit 0 from the first cache location will move into the bit 0 position in the second cache location. Bit 1 will move to bit 1, bit 2 will move to bit 2, and so on. That is, the cache can be thought of as being partitioned into 1,024+ spatially independent caches in order to make the $x$–$y$ dislocation 1,024+ times less general. Another critical point to note is that a given congruence class at one level can only receive data from a small subset of the congruence classes (as determined by address) at the next highest level. So the cache hierarchy can be further partitioned spatially by congruence classes (or "sets"). For example, if there are 1,024 congruence classes at the lowest layer in the cache hierarchy, this further makes the $x$–$y$ dislocation less general by another factor of 1,024.

In summary, by paying close attention to a number of critical factors when planning a stacked cache hierarchy, it is possible to reduce the $x$–$y$ dislocations in the bus infrastructure significantly (by six orders of magnitude in this example) within a 3D stack. Naturally, the actual number depends on the line size and on the specific geometry (set-associativity and congruence class structure) of the cache. It is possible to design the vertical buses nearly purely vertical and wide for many 3D cache hierarchy implementations. Such structures will enable virtualization at lower levels in the software stack, since they would facilitate the swapping of entire contexts fairly quickly and at reasonably low power.

548

## Challenges

Several critical challenges must be surmounted to make 3D structures widely practicable, aside from the obvious manufacturing challenges. First, there are new issues in building, assembly, and test (BAT) processes required to produce a structure that can be tested adequately during various stages of assembly and provide good final yield. In particular, since the individual layers are subject to their own yield issues, it is best to have individually testable layers prior to integrating them so that they will be working with known-good components. This will likely require more subcomponent redundancy with fairly flexible reconfigurability.

Note also that from an assembly point of view, wafer-to-wafer bonding is more practical than chip-to-wafer or chip-to-chip bonding. Yet, enabling this requires higher levels of redundancy to achieve high-yielding wafers. If the layers are to be individually testable, this is most easily accomplished if each layer is (from some point of view) a working independent system, which certainly describes the second system composition scenario discussed above and arguably describes the third. Testing an individual device layer in the stack requires that it have signals that can be contacted with test probes. Unless the contact points have extremely large capacitance as for power and ground, they will require ESD (electrostatic discharge) protection circuits, which are very large relative to the via pitch. This implies that there cannot be many contact points (certainly not per via), which likely requires that all signals entering or leaving a stratum be accessible by boundary scan latches.

Another key question is how many layers should be in an optimal stack. If 2D scaling is no longer advantageous (note that we are not claiming that this is the case, although the scaling is certainly slowing), then for 3D technology to provide the future scaling of a Moore's Law, the number of layers may increase with each generation. Hence, while the first product to emerge may have only two layers providing a density increase approximately equal to that of a new technology node, a long-term strategic road map must show four layers, eight layers, and so on as being possible. However, what is the limit, and where does this kind of 3D roadmap end (assuming that it is viable in the first place)? We state as intuitively obvious that the stack should stop before it is higher than a cube, as the advantages are likely to saturate with such stacking (such as the interconnect performance [2]).

For example, let the area of a layer be $A = x^2$; that is, let the chip be $x$ long on each edge. Assume that we have a stack of $n$ layers. Does it make more sense to add another layer ($n + 1$), or does it make more sense to increase the size of a layer to $x + \delta$? Adding an $n + 1$st layer will increase the total area of the structure that is usable for circuits by $A = x^2$. Increasing the area of the original structure to $(x + \delta)^2$ will increase the total area of the stack to $n(x^2 + 2\delta x + \delta^2)$. For the question to be at all interesting, we assume that $n$ is large enough so that $\delta$ is small enough to make $n\delta^2$ insignificant. So the total area added by increasing the areal footprint is essentially $2n\delta x$, compared to the $x^2$ increase gained by adding a layer.

This means that if $n < x/2\delta$, we get more usable circuit area by adding a layer; otherwise, it makes more sense to increase the footprint. Note that there is an interesting symmetry between $n$ and $\delta$, since we could also say that if $\delta < x/2n$, it makes more sense to add a layer than to increase the footprint.

Layer thickness is another important factor since it affects the impedance of a TSV running through the entire stack as well as the total height of the stack. Hence, a critical question is how thin the layers can be fabricated. Superficially, this appears to be purely a processing and yield question, but if we are considering thicknesses of a few microns, it also becomes an electrical question. In particular, if the bulk of a layer becomes too thin, can ground loops form where there were none before?

In the arguments mentioned above, we discussed silicon area as if it were all usable, but the usable area is at most the area that is not occupied by TSVs. As a stack gets progressively higher, it will require more power, hence more vias for power delivery. In fact in the limit (as $n$ becomes arbitrarily large), we would likely need 100% of the area for power vias. Obviously, such a system makes no sense, but it is interesting to consider this limit. We posit (but do not prove here) that the point of inflection for the fraction of the area used for TSVs in an optimal case is $1/e$, or 0.37. If more TSVs than this are required, the stack is likely too high from an optimization point of view (for given maximum power dissipation and TSV current limitations).

Just as power has become a limitation of 2D scaling, it will be in the forefront as a limitation in 3D scaling, and its effects will be more immediate. The challenge is twofold: One is power delivery as described above, and the other is heat removal. It is important to note that 3D technology facilitates much denser integration and thereby creates a power density problem immediately, particularly for logic stacking alternatives with high power density in 2D components. Lower total system power has always been the benefit of improved density, even though it inevitably leads to a power density problem as it scales. It is a wonderful conundrum to have.

While we have discussed, albeit indirectly, power delivery in proxy form as via area and we have posited a limit of $1/e$ of the area, there are several other aspects of power delivery that are interesting. One is to have 3D voltage-regulating layers that allow local voltage conversion and modulation to aid in power and thermal management. This will also facilitate power distribution

P. G. EMMA AND E. KURSUN

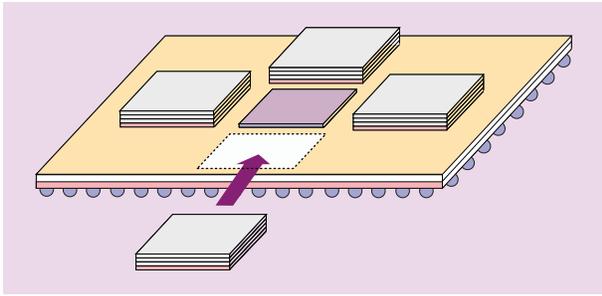

**Figure 7**

Stacks on a carrier—3½ dimensional?

at higher voltages within the system, thereby making currents more manageable.

The thermal profile of the 3D stack, especially in the case of logic–logic stacking, appears to be a major limitation of the vertical integration technology. As the packaging density is increased with the additional number of layers on a smaller footprint, the thermal profile is considerably increased. Furthermore, the thermal gradient problems are exacerbated because of the unique characteristics of the 3D stack. In the case of using traditional cooling solutions at the periphery of the stack, the thermal resistances accumulate vertically on the heat removal path. According to recent studies, the oxide material in the back-end-of-line (BEOL) wiring levels has been shown to have significantly high thermal resistance [7]. Hence, thermal-aware placement of functionality on individual layers has become an important consideration for 3D integration. The thermal characteristics of the 3D stack can exacerbate reliability issues, which in turn can pose a threat to system functionality if the hot areas are not managed. Increased temperature is likely to intensify electromigration in vertical signal interconnections and to cause device failures and aging.

Thermal vias and redundant power/ground vias have been shown to improve the thermal conductivity of 3D stacks [35]. However, the potential thermal improvement for aggressive stacking with high power density may not be sufficient to meet the thermal operation requirements. The need to invest in 3D-specific thermal management and cooling technologies [36] in order to cope with the total power density and corresponding thermal problems is becoming clearer, especially beyond the preliminary memory stacking implementations.

## Future 3D systems

It is fascinating to ponder the kinds of systems that could be possible to build if the challenges discussed above are surmounted and if the advantages discussed above are realized. These are systems that are beyond the scale of anything possible today. Ironically, the question is whether some of these ultimate systems are actually useful, and if so, then for what applications? It is likely that these ultimate systems will facilitate new applications that are not fathomable today.

One interesting observation is that silicon carriers and 3D stacks were once viewed as competing technologies that delivered approximately the same benefits at different processing costs. On reflection, one should be able to see that in fact, they are synergistic technologies. We can use silicon carriers to integrate an array of stacks into a socketable entity, as shown in **Figure 7**. A carrier of stacks (having a single connecting footprint) is essentially a 3½-dimensional system. Extrapolating the technology nodes in it down to the 32- or 22-nm nodes could enable several hundred cores and potentially hundreds of gigabytes to be integrated in a single system. Within that structure, we could have thousands of coincident threads, as well as the ability to move hundreds of pages around very quickly. What would we use this for? We believe that structures such as this would cause much of the software stack to be rethought.

Much of the complexity in the software stack exists because the system needs to manage many simultaneous processes exactly for the reason that those processes cannot access memory quickly enough, and the memory cannot deliver sufficient bandwidth to those processes. Three-dimensional structures have the potential to remove many of the bottlenecks that have been thought to be fundamental since the time of von Neumann.

The great irony is that the "doom and gloom" of the impending "end of Moore's Law" scenario presages a revolution that will likely foster innovation in software and microarchitecture at a scale that makes our last 20 years seem tame. In fact, as an industry, Moore's Law has allowed some amount of complacency to set in. We are about to mix things up!

**551**

**Philip G. Emma**  *IBM Research Division, Thomas J. Watson Research Center, P.O. Box 218, Yorktown Heights, New York 10598 (pemma@us.ibm.com)*. Dr. Emma received B.S., M.S., and Ph.D. degrees in electrical engineering from the University of Illinois, joining the IBM T. J. Watson Research Center in 1983. Since then, he has done extensive work in the areas of computer microarchitecture, computer architecture, systems design, circuit design, electronic packaging, materials, interconnection technology, and optics. He led the definition and design of the R-unit on the first IBM CMOS processors for IBM zSeries® machines and was responsible for the reliability and infrastructure that makes the zSeries machines the premier brand for reliable computing in the industry. Dr. Emma has written more than 200 technical articles and several book chapters, he holds 35 Invention Plateaus and more than 100 patents, and he has received several corporate awards for his technical work. He is an IBM Master Inventor, a member of the IBM Academy of Technology, and a Fellow of the Institute of Electrical and Electronics Engineers. He is currently Manager of the IBM Systems Technology and Microarchitecture Department.

**Eren Kursun**  *IBM Research Division, Thomas J. Watson Research Center, P.O. Box 218, Yorktown Heights, New York 10598 (ekursun@us.ibm.com)*. Dr. Kursun received her B.S. degree in electrical engineering from Bogazici University, Istanbul, in 2000. She received M.S. and Ph.D. degrees in computer science from the University of California, Los Angeles, in 2003 and 2006, respectively. She is currently a Research Staff Member in the Reliability and Power-Aware Microarchitectures groups. Her current research projects are in technology-aware microprocessor design and power and temperature management for microprocessor architectures.

**552**